# Stripe-like quasi-nondiffracting optical lattices

Yaroslav V. Kartashov,[1] Servando López-Aguayo,[2] Victor A. Vysloukh,[1] Lluis Torner[1]

[1]*ICFO- Institut de Ciencies Fotoniques, and Universitat Politecnica de Catalunya, Mediterranean Technology Park, 08860, Castelldefels (Barcelona), Spain*

[2]*Photonics and Mathematical Optics Group, Tecnologico de Monterrey, Monterrey, México 64849*

*Yaroslav.Kartashov@icfo.es*

**Abstract:** We introduce stripe-like quasi-nondiffracting lattices that can be generated via spatial spectrum engineering. The complexity of the spatial shapes of such lattices and the distance of their almost diffractionless propagation depend on the width of their ring-like spatial spectrum. Stripe-like lattices are extended in one direction and are localized in the orthogonal one, thereby creating either straight or curved in any desired fashion optically-induced channels that may be used for optical trapping, optical manipulation, or optical lattices for quantum and nonlinear optics applications. As an illustrative example, here we show their potential for spatial soliton control. Complex networks consisting of several intersecting or joining stripe-like lattices suited to a particular application may also be constructed.

**OCIS codes:** (260.1960) Diffraction theory; (190.6135) Spatial solitons

Nondiffracting light beams find important applications in many and diverse areas, that go from atom optics and nonlinear optics to microscopy and optical tweezing and trapping [1]. In many applications, the complexity of the transverse intensity distributions is as important as the diffractionless propagation. However, patterns used to date are limited to the known sets of simple nondiffracting beams that are exact solutions of the Helmholtz equation. Namely, rigorous group theory shows that there are only four different coordinate systems where the Helmholtz equation is separable, yielding corresponding invariant solutions along the propagation axis: plane waves in Cartesian coordinates, Bessel beams in circular cylindrical coordinates [2], Mathieu beams in elliptic cylindrical coordinates [3], and parabolic beams in parabolic cylindrical coordinates [4]. Therefore, construction of more sophisticated nondiffracting, or at least slowly diffracting, light beams with new unusual symmetries is important in view of applications that require specific shapes which do not belong to the above mentioned set.

In this work we introduce the possibility to generate stripe-like quasi-nondiffracting lattices, using spatial spectrum engineering techniques [5] - a method that is reminiscent to approaches used in phase retrieval and image processing algorithms [6]. By changing the spatial spectrum width one can generate patterns with stripe-like shapes that are almost nondiffracting over the depth of focus of interest, as long as the spatial spectral width is sufficiently small. Note that in practice the new patterns may be utilized in optical induction techniques [7-10], thus allowing the formation of the corresponding reconfigurable quasi-non-diffracting landscapes. To illustrate the usefulness of the new non-diffracting patterns, we show their application to spatial soliton control. Spatial solitons supported by such lattices posses unusual properties in comparison with their counterparts supported by uniform nonlinear media [11,12]. To date, the analysis of the properties of solitons in nondiffracting lattices has been restricted to the case of patterns produced by the above mentioned harmonic, Bessel, Mathieu, and parabolic beams [8,13-18]. However, we stress that our results are intended to have applications in all areas where complex optical lattices may be of interest.

By and large, nondiffracting light beams propagating along the $\xi$ axis may be written in terms of the Whittaker integral:

$$q_{\mathrm{nd}}(\eta,\zeta,\xi) = \exp(-ik_\mathrm{t}^2\xi/2)\int_0^{2\pi} G(\varphi)\exp[ik_\mathrm{t}(\eta\cos\varphi + \zeta\sin\varphi)]d\varphi, \qquad (1)$$

where $k_\mathrm{t} = (k_\eta^2 + k_\zeta^2)^{1/2}$ is the transverse wavenumber, $\varphi$ is the azimuthal angle in the frequency space, $\eta,\zeta$ are the transverse coordinates, and $G(\varphi)$ is the angular spectrum. The Fourier transform of $q_{\mathrm{nd}}$ reveals that the angular spectrum $G(\varphi)$ of the nondiffracting beam is defined on an infinitely narrow ring of the radius $k_\mathrm{t}$. Spatial spectrum broadening allows generation of patterns with more sophisticated shapes, since the spectral components with different transverse wavenumbers $k_\mathrm{t}$ from the finite band $\delta k_\mathrm{t}$ contribute to the beam profile. Such beams will diffract upon propagation due to dephasing of spectral components. However if the spectrum width is small $\delta k_\mathrm{t} \ll k_\mathrm{t}$, the diffraction may be correspondingly slow.

In order to generate quasi-nondiffracting stripe-like patterns, we first select the desired field distribution $\tilde{q}(\eta,\zeta)$ at $\xi=0$, where $|\tilde{q}(\eta,\zeta)|$ defines the fixed intensity distribution, while the phase profile $\arg[\tilde{q}(\eta,\zeta)]$ is a free parameter. After calculation of the Fourier transform $\tilde{q}(k_\eta,k_\zeta)$, we set to zero all components of the angular spectrum falling outside the annular ring of the width $\delta k_\mathrm{t}$ and radius $k_\mathrm{t}$. After this step, an inverse Fourier transform is calculated and the field modulus is replaced with the desirable $|\tilde{q}(\eta,\zeta)|$ distribution, while keeping the new phase distribution. This iterative procedure is repeated until the convergence is achieved. We use the field from last iteration without replacing its modulus with $|\tilde{q}(\eta,\zeta)|$. While for large relative width of the angular spectrum $\delta = \delta k_\mathrm{t}/k_\mathrm{t}$ the resulting beam can accurately reproduce almost any desirable initial pattern $|\tilde{q}(\eta,\zeta)|$, for $\delta \ll 1$ the iterative procedure always introduces distortions into the initial "ideal" distribution. Further we set $k_\mathrm{t} = 2^{1/2}$ and $\delta = 0.1$ to ensure that our patterns will propagate without noticeable diffraction over more than twenty diffraction lengths.

Examples of stripe-like patterns are presented in Fig. 1. Remarkably, such patterns are extended in one direction and are strongly localized in the orthogonal one [Fig. 1(a)]. In the case of a single stripe the spatial spectrum looks like a straight segment, indicating that the beam forms due to the interference of plane waves with almost identical $k_\eta$ and different $k_\zeta$ wavenumber components [Fig. 1(b), left]. It is possible to join stripe-like beams in one point [Fig. 1(a), center] or intersect them [Fig. 1(a), right], creating star-like patterns. Interestingly, the pattern with three branches [Fig. 1(a), center] has a much richer spatial spectrum. The intersection or fusion of several stripes results in a small-scale interference pattern in the crossing region: the interference may be destructive [Fig. 1(a), center] or constructive [Fig. 1(a), right]. Notice that different stripes can have different amplitudes [see Fig. 1(c) where we show a gradual transformation between single- and double-stripe structures upon increase of the amplitude of vertical stripe]. Finally, a superposition of several shifted stripes produces really complex patterns [Fig. 1(c), right].

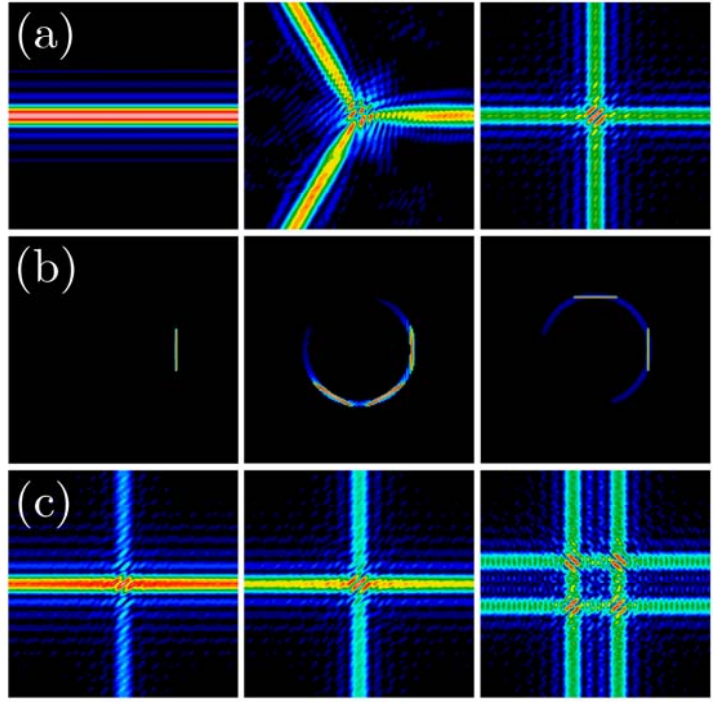

Fig. 1. (a) Field modulus distributions and (b) spatial spectra for star-like quasi-nondiffracting beams with two, three, and four branches, respectively. Row (c) shows quasi-nondiffracting beams with gradually increasing amplitudes of vertical stripes and complex beam produced by the intersection of two vertical and two horizontal stripes. In all cases $\delta = 0.1$.

The stripe-like patterns may be used for optical induction of guiding channels where spatial solitons may form, travel, and interact. The nonlinear propagation of light beams in such optically induced lattices can be described by the equation:

$$i\frac{\partial q}{\partial \xi} = -\frac{1}{2}\left(\frac{\partial^2 q}{\partial \eta^2} + \frac{\partial^2 q}{\partial \zeta^2}\right) - Eq\frac{|q|^2 + pR}{1+|q|^2 + pR}, \qquad (2)$$

where we used a simplified isotropic model of a photorefractive crystal response [7-10], the parameter $E$ is proportional to the external biasing field, the lattice shape is described by the function $R \sim |q_{\mathrm{nd}}|^2$ that is proportional to intensity of nondiffracting beam, the parameter $p$ determines the lattice depth. Further we set $E = 10$. The soliton solutions of Eq. (2) have the form $q(\eta, \zeta, \xi) = w(\eta, \zeta)\exp(ib\xi)$, where $b$ is the propagation constant. Representative examples of solitons supported by lattices consisting of two intersecting stripe-like beams are shown in Fig.

2(a). The shapes of the solitons are strongly affected by the details of the interference pattern in the crossing region. In a single-stripe lattice [like the one shown in Fig. 1(a), left] that does not feature any refractive index modulation in the horizontal direction, low intensity solitons are strongly elongated along such direction (note that this is due to the fact that single-stripe lattices does not support any localized linear modes and this is in contrast to solitons in parabolic lattices addressed in [18] that bifurcate from localized linear guided modes of the lattice). However, when the amplitude of the vertical stripe increases, the interference pattern that appears in the crossing region causes a gradual reorientation and localization of the solitons even in low-power regime.

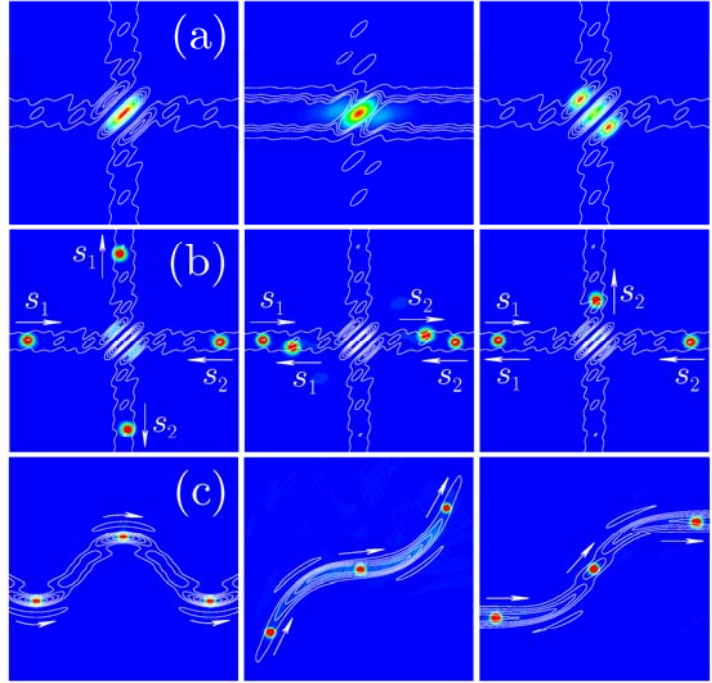

Figure 2. (a) Left and middle panels show fundamental solitons with $U=1$ supported by star-like lattices where relative amplitudes of vertical stripes are $1.0$ and $0.3$, respectively. Right panel shows field modulus distribution in tripole soliton with $U=27$ in star-like lattice where relative amplitude of vertical stripe is $1.0$. All solitons in (a) are obtained at $p=4$. Row (b) illustrates a variety of routing scenarios available in star-like lattices. The initial velocities of out-of-phase solitons "$s_1$", "$s_2$" are $\alpha_1=1.5$, $\alpha_2=-1.5$ (left), $\alpha_1=2.2$, $\alpha_2=-2.2$ (middle), and $\alpha_1=1.8$, $\alpha_2=-2.0$ (right). The input solitons correspond to $U=1.5$ and $p=0.2$. (c) Snapshot images showing propagation of solitons at $\alpha=1.6$, $b=4$, $p=0.5$ in bent lattices where guiding channels are periodically curved or follow cubic parabola or tanh curves. Arrows in (b) and (c) show the direction of soliton motion.

At low intensities the solitons are strongly elliptical due to the specific details of the interference pattern, but at high intensities they become almost circularly symmetric due to the nonlinearity saturation. In addition to fundamental states [Fig. 2(a), left and central panels] such lattices support dipole, tripole [Fig. 2(a), right], and higher-order soliton states. The energy flow $U$ [see Eq. (3) for definition] is a monotonically increasing function of the propagation constant [Fig. 3(a)], while the integral soliton width $W$, defined as

$$W = 2U^{-1} \int\int_{-\infty}^{\infty} (\eta^2 + \zeta^2)^{1/2} |q|^2 \, d\eta d\zeta, \tag{3}$$
$$U = \int\int_{-\infty}^{\infty} |q|^2 \, d\eta d\zeta,$$

diverges in the high-intensity limit and approaches a constant value for low-intensity solitons [Fig. 3(b)]. Solitons in such lattices exist between the lower $b_{\text{low}}(p)$ and upper $b_{\text{upp}} \equiv E$ cutoffs on the propagation constant. The domain of their existence $[b_{\text{low}}, b_{\text{upp}}]$ shrinks with increase of the lattice depth $p$ [Fig. 3(c)]. Both fundamental and multipole solitons are stable inside most of their existence domain.

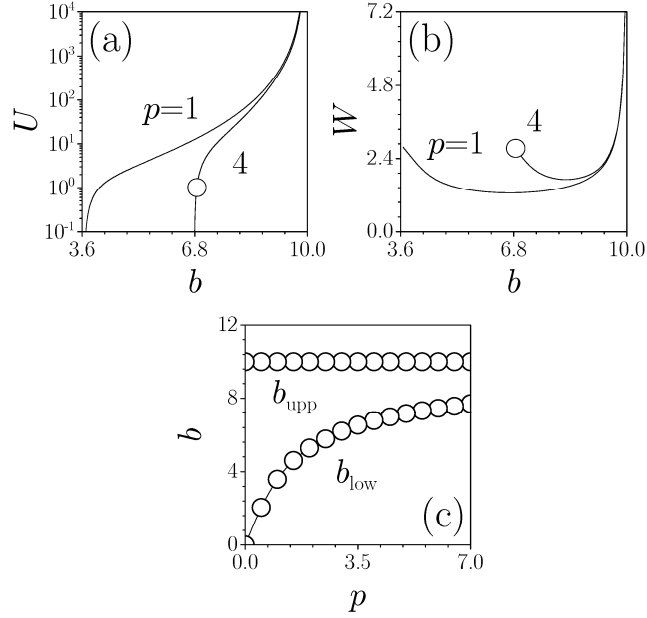

Fig. 3. Energy flow $U$ (a) and width $W$ (b) of fundamental soliton supported by the lattice with four branches versus propagation constant $b$. The point marked by circle corresponds to soliton in Fig. 2(a), left panel. (c) Upper and lower cutoffs versus lattice depth.

Stripe-like lattices offer new opportunities for soliton routing and control, because solitons can move almost freely along the channels and interact there. Thus, the result of interaction between out-of-phase solitons, launched initially in the opposite branches of the lattice produced by two intersecting stripes, depends drastically on the initial soliton velocities $\alpha_{1,2}$ [i.e. the phase tilts $\exp(i\alpha_{1,2}\eta)$ imprinted on the input solitons]. In a uniform nonlinear medium, such solitons after a head-on collision usually bounce back. In the lattice, featuring a complex refractive index profile in the intersection region, different scenarios are possible depending on the $\alpha_{1,2}$ values. The solitons can switch into the orthogonal lattice branches [Fig. 2(b), left], they can both bounce back after the interaction [Fig. 2(b), center], or one soliton may switch to the orthogonal branch, while the other one bounces back [Fig. 2(b), right].

Importantly, the channels of stripe-like lattices may not be necessarily straight. It is possible to form lattices with bent channels, which may follow the profiles of different functions in the $(\eta, \zeta)$ plane. Figure 4 shows examples of such lattices whose channels experience periodic bending in accordance with $\zeta \sim \cos(C\eta)$ law [Fig. 4(a)], follow the profile of a cubic parabola $\zeta \sim C\eta^3$ [Fig. 4(b)], or hyperbolic tangent function $\zeta \sim \tanh(C\eta)$ [Fig. 4(c)]. As one can see, the stronger the deviation of the lattice channel from the straight-line structure [like the one shown in Fig. 1(a)], the more pronounced the background and secondary intensity maxima sur-rounding the main channel. Solitons walk along a complex trajectory if launched into a bent lattice with an appropriate tilt in the direction tangential to the lattice channel [see Fig. 2(c) where snapshot images illustrate the soliton propagation paths for all lattices from Fig. 4].

It should be stressed that the stripe-like quasi-nondiffracting lattices addressed here may be advantageous in comparison with technologically fabricated waveguide arrays, because the refractive index modulation in such lattices is created all-optically and therefore it can be easily erased by additional illumination at specific wavelength or reconfigured by blocking several stripe-like beams or by launching additional beams into the medium. In practice, this can be achieved by introducing the corresponding modifications into the computer-generated holograms used to synthesize the beams experimentally. In this way the shape and topology of the lattice can be changed in real time, a property not possible in fabricated arrays.

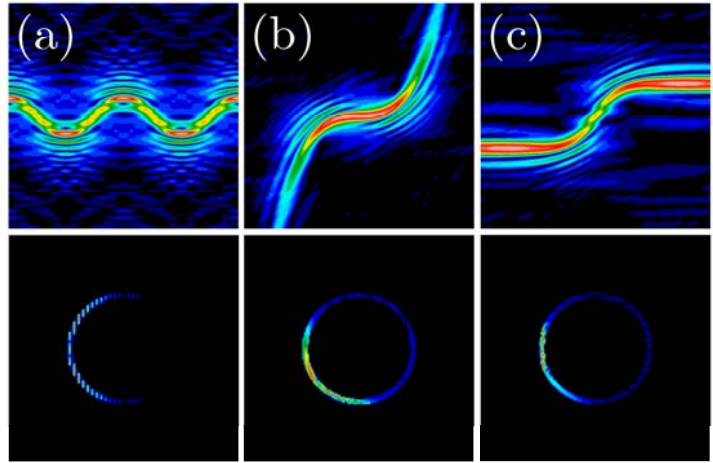

Fig. 4. Field modulus distributions (top) and spatial spectra (bottom) in stripe-like quasi-nondiffracting beams that bent periodically (a), follow the shape of cubic parabola (b) or hyperbolic tangent (c) functions. The spectrum width is $\delta = 0.1$ in all cases.

Summarizing we introduced a new type of quasi-nondiffracting optical lattice which exhibit strip-like shapes. For illustrative purposes, here we showed that such lattices may be utilized in spatial soliton routing schemes for implementation of switching operations, since the lattices feature pronounced straight or curved guiding channels where spatial solitons can move and interact with each other in a controlled and reconfigurable way. However, we stress that this is only one particular example of the potential application of such patterns, which may be useful in diverse areas beyond nonlinear optics where the corresponding reconfigurable, complex optical lattices may be important, such us optical manipulation and transport of living material in biophysics or of single atoms and BECs in quantum information operations.